Chapter 1

# Internet of Things (IoT) and New Computing Paradigms

*Chii Chang, Satish Narayana Srirama and Rajkumar Buyya*

## 1.1 Introduction

The Internet of Things (IoT) [1] represents a comprehensive environment that interconnects a large number of heterogeneous physical objects or things such as appliances, facilities, animals, vehicles, farms, factories etc. to the Internet, in order to enhance the efficiency of the applications such as logistics, manufacturing, agriculture, urban computing, home automation, ambient assisted living and various real-time ubiquitous computing applications.

Commonly, an IoT system follows the architecture of the Cloud-centric Internet of Things (CIoT) in which the physical objects are represented in the form of Web resources that are managed by the servers in the global Internet [2] . Fundamentally, in order to interconnect the physical entities to the Internet, the system will utilize various front-end devices such as wired or wireless sensors, actuators, readers to interact with them. Further, the front-end devices have the Internet connectivity via the mediate gateway nodes such as Internet modems, routers, switches, cellular base stations and so on. In general, the common IoT system involves three major technologies: embedded systems, middleware and cloud services, where the embedded systems provide intelligence to the front-end devices, middleware interconnects the heterogeneous embedded systems of front-end devices to the cloud and finally, the cloud provides comprehensive storage, processing, and management mechanisms.

Although the CIoT model is a common approach to implement IoT systems, it is facing the growing challenges in IoT. Specifically, CIoT faces challenges in BLURS—Bandwidth, Latency, Uninterrupted, Resource-constraint and Security [3] .

- **Bandwidth**. The increasingly large and high-frequent rate data produced by objects in IoT will exceed the bandwidth availability. For example, a connected car can generate tens of megabytes' data per second for the information of its route, speeds, car operating condition, driver's condition, surrounding environment, weather etc. Further, a self-driving vehicle can generate gigabytes of data per second due to the need for real-time video streaming. Therefore, fully relying on the distant cloud to manage the things becomes impractical.

- **Latency**. Cloud faces the challenges to achieve the requirement of controlling the end-to-end latency within tens of milliseconds. Specifically, industrial smart grids systems, self-driving vehicular networks, virtual and augmented reality applications, real-time financial trading applications, healthcare and eldercare applications cannot afford the causes derived from the latency of CIoT.

- **Uninterrupted**. The long distance between cloud and the front-end IoT devices can face issues derived from the unstable and intermittent network connectivity. For example, a CIoT-based connected vehicle will be unable to function properly due to the disconnection occurred at the intermediate node between the vehicle and the distant cloud.

- **Resource-Constrained**. Commonly, many front-end devices are resource-constraint in which they are unable to perform complex computational tasks and hence, CIoT systems usually require front-end devices to continuously stream their data to the cloud. However, such a design is impractical in many devices that operate with battery power because the end-to-end data transmission via the Internet can still consume a lot of energy.

- **Security**. A large number of constraint front-end devices may not have sufficient resources to protect themselves from the attacks. Specifically, outdoor-based front-end devices, which rely on the distant cloud to keep them updated with the security software, can be attackers' targets, in which the attackers are capable of performing a malicious activity at the edge network where the front-end devices are located and the cloud does not have full control on it. Furthermore, the attacker may also damage or control the front-end device and send false data to the cloud.

The growing challenges of CIoT raised a question—*what can be done to overcome the limitation of current cloud-centric architecture?*

In the last decade, several approaches have tried to extend the centralized cloud computing to a more geo-distributed manner in which the computational, networking and storage resources can be distributed to the locations that are much closer to the data sources or end-user applications. For example, the geo-distributed cloud computing model [4] tends to partition the portions of processes to the data centers near the edge network. Further, mobile cloud computing model [5] introduced the physical proximity-based cloud computing resources provisioned by the local wireless Internet access point providers. Moreover, academic research projects [6] have experimented the feasibility of the mobile ad hoc network (MANET)-based cloud using the Advanced RISC Machine (ARM)-powered devices. Among the various approaches, the industry-led fog computing architecture, which was first introduced by Cisco research [7] has gained the most attention.

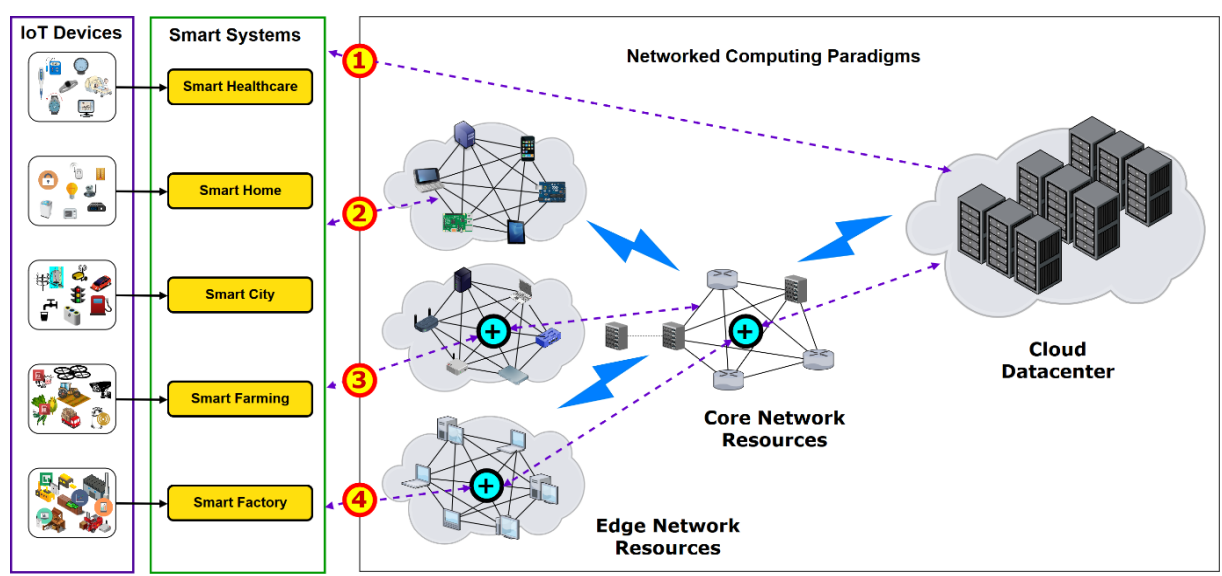

Figure 1.1: IoT applications and environments with supporting computing paradigms.

Fog computing architecture [8] covers a broad range of equipment and networks. In general, it is a conceptual model that address all the possibilities to extend the cloud to the edge network of CIoT, from the geo-distributed data center, intermediate network nodes to the extreme edge where the front-end IoT devices are located. Figure 1.1 illustrates different network computing paradigms supporting IoT-enabled smart systems and applications. To enumerate, general CIoT paradigm (mark 1) manages the smart systems entirely at the distant central cloud datacentre in which the IoT devices act as simple sensory data collectors or actuators, and leave the processes and decision-making to the cloud; generic edge computing paradigm (mark 2) distributes certain tasks to the IoT devices or the co-located computers within the same subnet of the IoT devices. Such tasks can be data classification, filtering, the signal converting, etc.; fog computing paradigm (marks 3 & 4) utilizes hierarchical-based distributed computing model which supports horizontal scalability of the computational resources. For example, a fog-enabled IoT system can distribute the simple data classification tasks to the IoT devices and assign the more complicated context reasoning tasks at the edge gateway devices. Further, for the analytics tasks that involve terabytes data, which requires higher processing power, the system can further move the processes to the resources at the core network such as the data centers of Wide Area Network (WAN) service providers or utilize the cloud. Certainly, the decision of where the system should assign the tasks among the resources across different tiers depends on the efficiency and adaptability. For example, smart systems may need to assign certain decision-making tasks to the edge devices in order to provide timely notification about the situation such as the patient's condition in the smart healthcare, the security state of the smart home, the traffic condition of the smart city, the water supply condition of smart farming or the production line operation condition of smart factory.

The industry has seen fog as the main trend for the practical IoT systems and the leading OpenFog consortium has established collaboration with major industrial standard parties such as European Telecommunications Standards Institute (ETSI) Multi-access Edge Computing (MEC) and IEEE Standard for Fog Computing and Networking [9] to hasten the fog. Furthermore, the fog market research report [10] stated that the market value of fog will grow from $3.7bn by 2019 up to $18.2bn by 2022 across different fields, where the top five utilization domains of fog will be energy/utilities, transportation, healthcare, industrial and agriculture.

In this chapter, we discuss foundations of computing paradigms for realizing emerging IoT applications, especially fog and edge computing, their background, characteristics, architectures and open challenges. Section 1.2 presents related technologies to fog and edge computing. Section 1.3 describes how fog and edge can improve CIoT. Section 1.4 explains the hierarchy of fog and edge computing environments. Section 1.5 illustrates the business models of fog and edge computing. Section 1.6 provides the information regarding to the opportunities and challenges in fog and edge computing. Finally, Section 1.7 summarizes the content of the Chapter.

## 1.2   Relevant Technologies

The notion of having computational resources near the data sources may seem not new. Particularly, the term—edge computing appeared in 2004 to illustrate a system that distributes program methods and the corresponding data to the network edge towards enhancing performance and efficiency [11] . Similarly, the notion of having virtualization technology-based computing resources within the Wi-Fi subnet has been introduced in 2009 [5]  However, the real industrial interest in extending computational resources to the edge network only started after the introduction of fog computing for IoT. Prior to that, applying utility cloud at the edge network was more or less a research topic in academia without explicit definition, architecture and with minor industrial involvement. In contrast, the industry has invested fog computing architecture by establishing OpenFog consortium founded by ARM Holdings, Cisco, Dell, Intel, Microsoft, Princeton University and over 60 members from major industrial and academic partners in the world. Further, in collaboration with international standard organizations such as ETSI and IEEE, fog has become a major trend in general Information and Communication Technology (ICT) today.

In last several years, researchers have been using different terminologies to illustrate the similar architectures with fog. For example, the author of Virtual Machine (VM)-based cloudlet [5] tended to use edge computing to describe the notion of cloud at the edge. Moreover, the author's later work indicated that fog is a part of edge computing [12] . On the other hand, OpenFog consortium has specifically differentiated the two terminologies. Explicitly, the initial objective of cloudlet was to provide the mobile application a substitution from the distant cloud, in which the mobile applications can offload computing-intensive tasks to the nearby cloudlet VM machines co-located within the same Wi-Fi subnet. Whereas, the initial introduction of fog computing aimed to complete the cloud by extending the cloud to the network gateways themselves. In essence, cloudlet can be seen as one of the practical approaches for fog computing when the co-located physical server machines are available.

Certain other works have been describing Multi-access Edge Computing (MEC; formerly Mobile Edge Computing) as an exchangeable term with fog. Essentially, ETSI introduced MEC as a standard from the perspective of telecommunication, in which ETSI specifies the Application Programming Interface (API) standards about how telecommunication companies can provide computing virtualisation-based service to their clients based on extending the existing infrastructure used in Network Function Virtualisation (NFV), which has been already implemented in existing equipment such as cellular Base Transceiver Stations (BTSs). Although it is inaccurate to describe MEC as an exchangeable term as fog, according to the recent collaboration between OpenFog and ETSI, MEC will become a practical approach to hasten the realization of fog computing [13] .

Mist computing was an alternative term for fog in the earlier stages. However, recent works have described mist as a subset of fog. Accordingly, mist elaborates the need of distributing computing mechanism to the extreme edge of IoT, where the IoT devices are located, in order to minimize the communication latency between IoT devices in milliseconds [14] [15] [16] . Essentially, the motivation of mist computing is to grant the IoT devices with the capability of self-awareness in terms of self-organizing, self-managing and several self-* mechanisms. Therefore, the IoT devices will be able to continuously operate even when the Internet connection is unstable.

In general, mist devices may sound similar to the embedded services or mobile Web services [17] in which the application services are hosted in heterogeneous resource-constrained devices such as sensors, actuators, and mobile phones. However, mist emphasizes the capability of self-awareness and situation-awareness in which it allows dynamic and remotely (re)deploying software program code to the devices based on the situation and context changes [14] . Such a feature shares similarity with fog in providing a platform that allows flexible software deployment and reconfigurations.

Realizing that, the fog requires the support of all the related edge computing technologies. In other words, one is unable to deploy and manage fog without integrating edge computing technologies. Therefore, in the rest of this chapter, we use the term—Fog and Edge Computing (FEC) to describe the whole domain.

## 1.3 Fog and Edge Computing Completing the Cloud

FEC provides a complement to the cloud in IoT by filling the gap between cloud and things towards providing service continuum [3] . In particular, FEC offers five main advantages, which can be exemplified by SCALE—Security, Cognition, Agility, Latency, and Efficiency [8] .

*Security.* FEC supports additional security to IoT devices to ensure safety and trustworthiness in transactions. For example, today's wireless sensors deployed in outdoor environments often require the remote wireless source code update in order to resolve the security-related issues. However, due to various dynamic environmental factors such as unstable signal strength, interruptions, constraint bandwidth etc., the distant central backend server may face challenges to perform the update swiftly and hence, increases the chance of cyber security attack. On the other hand, if the FEC infrastructure is available, the backend can configure the best routing path among the entire network via various FEC nodes in order to rapidly perform the software security update to the wireless sensors.

***Cognition.*** FEC enables the awareness of the objectives of their clients towards supporting autonomous decision making in terms of where and when to deploy computing, storage and control functions. Essentially, the awareness of FEC, which involves a number of mechanisms in terms of self-adaptation, self-organization, self-healing, self-expression and so forth [16] , shifts the role of IoT devices from passive to active smart devices that can continuously operate and react to customer requirements without relying on the decision from the distant cloud.

***Agility.*** FEC enhances the agility of the large scope IoT system deployment. In contrast to the existing utility cloud service business model, which relies on the large business holder to establish, deploy and manage the fundamental infrastructure, FEC brings the opportunity to individual and small businesses to participate in providing FEC services using the common open software interfaces or open Software Development Kits (SDKs). For examples, the MEC standard of ETSI and the Indie Fog business model [18] will hasten the deployment of large scope IoT infrastructures.

***Latency.*** The common understanding of FEC is to provide rapid responses for the applications that require ultra-low latency. Specifically, in many ubiquitous applications and industrial automation, the system needs to collect and process the sensory data continuously in the form of the data stream in order to identify any event and to perform timely actions. Explicitly, by applying FEC, these systems are capable of supporting time-sensitive functions. Moreover, the softwarization feature of FEC, in which the behavior of physical devices can be fully configured by the distant central server using software abstraction, provides a highly flexible platform for rapid re-configuration of the IoT devices.

***Efficiency.*** FEC enhances the efficiency of CIoT in terms of improving performance and reducing the unnecessary costs. For example, by applying FEC, the ubiquitous healthcare or eldercare system can distribute a number of tasks to the Internet gateway devices of the healthcare sensors, and utilize the gateway devices to perform the sensory data analytics tasks. Ideally, since the process happens near the data source, the system can generate the result much faster. Further, since the system utilizes gateway devices to perform most of the tasks, it highly reduces the unnecessary cost of outgoing communication bandwidth.

The high-level description of the advantages provided by FEC leads to a question: *How does FEC provide these advantages?* To answer the question, here, we describe the five basic mechanisms supported by FEC-enabled devices (FEC node; see Figure 1.2). Specifically, the mechanisms can be termed as SCANC, which corresponds to Storage, Compute, Acceleration, Networking, and Control.

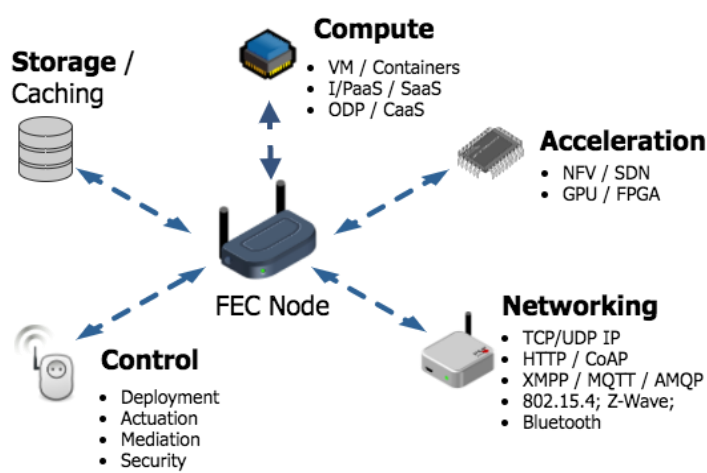

**Figure 1.2. FEC nodes supports five basic mechanisms—storage, compute, acceleration, networking and control.**

***Storage.*** The mechanism of storage in FEC corresponds to the temporary data storing and caching at the FEC nodes in order to improve the performance of information or content delivery. For example, content service providers can perform multimedia content caching at the FEC nodes that are most close to their customers in order to improve the quality of experience [19] . Further, in connected vehicle scenarios, the connected vehicles can utilize the roadside FEC nodes to fetch and to share the information collected by the vehicles continuously.

***Compute.*** FEC nodes provide the computing mechanisms mainly in two models—Infrastructure or Platform as a Service (I/PaaS) and Software as a Service (SaaS). In general, FEC providers offer I/PaaS based on two approaches—hypervisor Virtual Machines (VMs) or Containers Engines (CEs), which enable flexible platforms for FEC clients to deploy the customized software they need in a sandbox environment hosted in FEC nodes. Besides the I/PaaS, the SaaS is also promising in FEC service provision [3] . To enumerate, SaaS providers can offer two types of services—On-demand Data Processing (ODP) and Context as a Service (CaaS). Specifically, an ODP-based service has pre-installed methods that can process the data sent from the client in the request/response manner. Whereas, the CaaS-based service provides a customized data provision method in which the FEC nodes can collect and process the data to generate meaningful information for their clients.

***Acceleration.*** FEC provides acceleration with a key concept—"programmable". Fundamentally, FEC nodes support acceleration in two aspects—networking acceleration and computing acceleration.

- **Networking acceleration**. Initially, most network operators have their own configuration in message routing paths in which their clients are unable to request for the customized routing tables. For example, an Internet Service Provider (ISP) in East Europe may have two routing paths with different latency to reach a Web server located in Central Europe, and the path a client will be on is based on the ISP's load balancing setting, which in many cases, is not the optimal option for the client. On the other hand, FEC supports network acceleration mechanism based on network virtualization technology, which enables FEC nodes to operate multiple routing tables in parallel and to realize Software Defined Network (SDN). Therefore, the clients of the FEC nodes can configure customized routing path for their applications in order to achieve optimal network transmission speed.
- **Computing acceleration**. Researchers in fog computing have envisioned that the FEC nodes will provide computing acceleration by utilizing advanced embedded processing units such as Graphics Processing Units (GPUs) or Field Programmable Gate Arrays (FPGA) units [8] . Specifically, utilizing GPUs to enhance the process of complex algorithms has become a common approach in general cloud computing field. Therefore, it is foreseeable that FEC providers may also provide the equipment that contains middle- or high-performance independent GPUs. Further, FPGA units allow users to re-deploy program codes on them in order to improve or update the functions of the host devices. Particularly, researchers in sensor technologies [20] have been utilizing FPGA for runtime reconfiguration of sensors for quite some time. Further, in comparison with GPUs, FPGA is potential to be a more energy efficient approach for the need of acceleration based on allowing clients to configure their customized code on the FEC nodes.

***Networking.*** Networking of FEC involves vertical and horizontal connectivities. Vertical networking interconnects things and cloud with the IP networks; whereas, horizontal networking can be heterogeneous in network signals and protocols depending on the supported hardware specification of the FEC nodes.

- **Vertical networking**. FEC nodes enable vertical network using IP network-based standard protocols such as the request/response-based TCP/UDP sockets, HTTP, Internet Engineering Task Force (IETF)---Constraint Application Protocol (CoAP) or publish-subscribe-based Extensible Messaging and Presence Protocol (XMPP), OASIS---Advanced Message Queuing Protocol (AMQP; ISO/IEC 19464), Message Queue Telemetry Transport (MQTT; ISO/IEC PRF 20922) and so forth. Specifically, the IoT devices can operate server-side function (e.g. CoAP server) that allows FEC nodes, which act as the proxy of cloud, to collect data from them and then forward the data to the cloud. Further, FEC nodes can also operate as the message broker of publish-subscribe-based protocol that allows the IoT devices to publish data stream to the FEC nodes and enable the cloud backend to subscribe the data stream from the FEC nodes.

- **Horizontal networking**. Based on various optimization requirements such as energy efficiency or the network transmission efficiency, IoT systems are often using heterogeneous cost-efficient networking approaches. In particular, smart home, smart factories, connected vehicles are commonly utilizing Bluetooth, ZigBee (based on IEEE 802.15.4), Z-Wave on the IoT devices and connect them to an IP network gateway towards enabling the connectivity between the devices and the backend cloud. In general, the IP network gateway devices are the ideal entities to host FEC servers since they have the connectivity with the IoT devices in various signals. For example, the cloud can request an FEC server hosted in a connected car to communicate with the roadside IoT equipment using ZigBee in order to collect the environmental information needed for analyzing the real-time traffic situation.

***Control.*** The control mechanism supported by FEC consists of four basic types---deployment, actuation, mediation, and security.

- **Deployment control** allows clients to perform customizable software program deployment dynamically. Further, clients can configure FEC nodes to control which program the FEC node should execute and when it should execute it? Further, FEC providers can also provide a complete FEC network topology as a service that allows clients to move their program from one FEC node to another. Moreover, the clients may also control multiple FEC nodes to achieve the optimal performance for their applications.
- **Actuation control** represents the mechanism supported by the hardware specification and the connectivities between the FEC nodes and the connected devices. Specifically, instead of performing direct interaction between the cloud and the devices, the cloud can delegate certain decisions to FEC nodes to directly control the behavior of IoT devices.
- **Mediation control** corresponds to the capability of FEC in terms of interacting with external entities owned by different parties. In particular, the connected vehicles supported by different service providers can communicate with one another though they may not have a common protocol initially, with the softwarization feature of FEC node, the vehicles can have on-demand software update towards enhancing their interoperability.
- **Security control** is the basic requirement of FEC nodes that allows clients to control the authentication, authorization, identify and protection of the virtualized runtime environment operated on the FEC nodes.

## 1.4 Hierarchy of Fog and Edge Computing

In general, from the perspective of central cloud in the core network, CIoT systems can deploy FEC servers at three edge layers—inner-edge, middle-edge, and outer-edge (see Figure 1.3). Here, we summarize the characteristics of each layer.

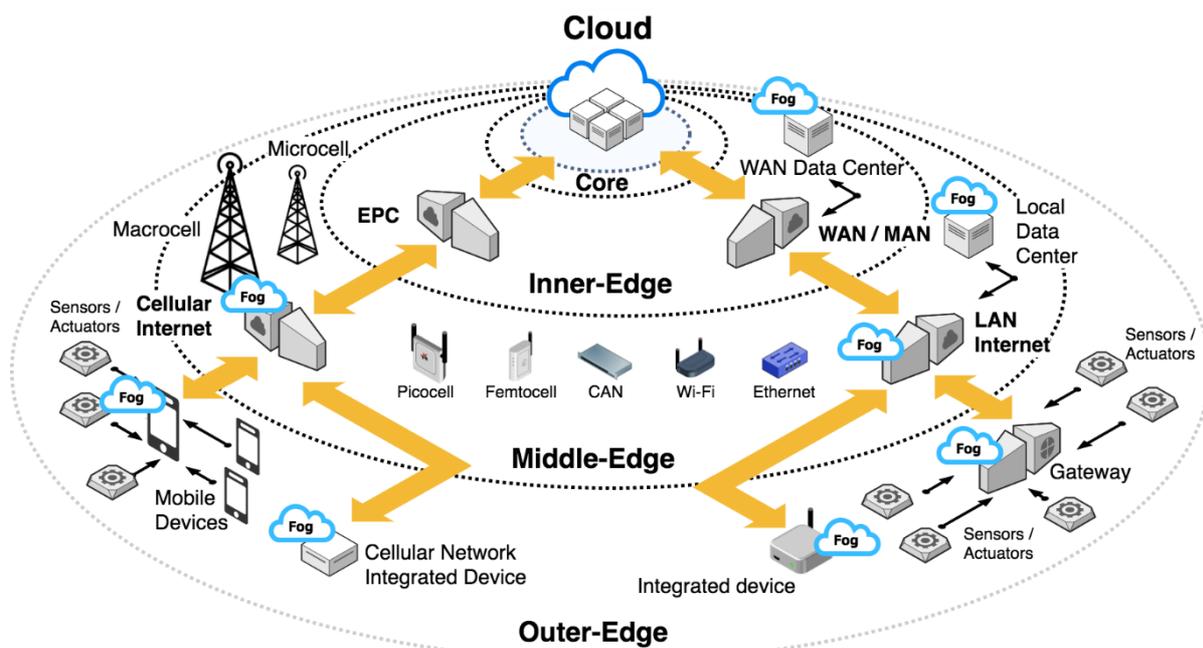

**Figure 1.3. Hierarchy of fog and edge computing.**

## *1.4.1 Inner-Edge*

*Inner-edge* (also known as near-the-edge [4] ) corresponds to countrywide, statewide and regional Wide Area Network (WAN) of enterprises, Internet Service Providers (ISPs), the data center of Evolved Packet Core (EPC) and Metropolitan Area Network (MAN). Initially, service providers at inner-edge only offer the fundamental infrastructures for connecting local networks to the global Internet. However, the recent needs in improving the Quality of Experience (QoE) of Web services have motivated the geo-distributed caching and processing mechanism at the network data centers of WAN. For example, in the commercial service aspect, Google Edge Network (peering.google.com) collaborates with ISPs to distribute data servers at the ISPs' data centers in order to improve the response speed of Google's cloud services. Further, many ISPs (e.g. AT&T, Telstra, Vodafone, Deutsche Telekom etc.) are aware that many local businesses require low latency cloud and hence, they have offered local cloud within the country. Based on the reference architecture of fog computing [8] , the WAN-based cloud data centers can be considered as the fog of inner-edge.

## *1.4.2 Middle-Edge*

*Middle-edge* corresponds to the environment of the most common understanding of FEC, which consists of two types of networks—Local Area Network (LAN) and cellular network. To summarize, LAN includes ethernet, Wireless LAN (WLAN) and Campus Area Network (CAN). Whereas, the cellular network consists of the macrocell, microcell, picocell, and femtocell. Explicitly, middle-edge covers a broad range of equipment to host FEC servers.

**Local Area Network.** The emerging fog computing architecture introduced by Cisco's research [7] was utilizing Internet gateway devices (e.g. Cisco IR829 Industrial Integrated Router) to provide the similar model as utility cloud services in which the gateway devices provide virtualization technologies that allow the gateway devices to support FEC mechanisms mentioned previously. Further, it is also an ideal solution to utilize the virtualization technology-enabled server computers located within the same subnet of LAN or CAN (i.e. within the one-hop range between the IoT device and the computer) with the FEC nodes. Ordinarily, such an approach is also known as local cloud, local data center or cloudlet.

**Cellular Network.** The idea of providing FEC mechanisms derived from the existing network virtualization technologies that have been used in various cellular networks. In general, most developed cities have wide coverage of cellular networks provided by numerous types of Base Transceiver Stations (BTSs), which are the ideal facilities to serve as roadside FEC hosts for various mobile IoT use cases such as connected vehicles, mobile healthcare, virtual or augmented reality, which require rapid process and response on the real-time data stream. Therefore, major telecommunication infrastructure and equipment providers such as Nokia, ADLink or Huawei have started providing MEC-enabled hardware and infrastructure solutions. Accordingly, it is foreseeable that in near future, cellular network-based FEC will be available in a broad range of related equipment from macrocell BTS, microcell BTS to the indoor cellular extension equipment such as picocell and femtocell [21] base stations.

## *1.4.3 Outer-Edge*

*Outer-edge*, which is also known as extreme-edge, far-edge or mist [14] [15] [16] , represents the front-end of the IoT network where consists of three types of devices—constraint devices, integrated devices and IP gateway devices.

**Constraint devices** such as sensors or actuators are usually operated by microcontrollers that have the very limited processing power and memory. For example, Atmel ATmega328 single-chip microcontroller, which is the CPU of Arduino Uno Rev3, has only 20 MHz processing power and 32kB flash memory. Commonly, IoT administrators would not expect to deploy complex tasks to this type of devices. However, due to the "field programmable" ability of today's wireless sensors and actuators, the IoT system can always update or re-configure the program code of the devices dynamically and remotely. Explicitly, such a mechanism grants the constraint IoT devices with self-awareness feature and motivated the mist computing discipline [14] , which emphasizes the abilities of IoT devices in self-management of the interaction and collaboration among IoT devices themselves towards achieving a highly autonomous Machine-to-Machine (M2M) environment without relying on the distant cloud for all their activities.

**Integrated devices** are the devices operated by the processors that have the decent processing power. Further, the integrated devices have many embedded capabilities in networking (e.g. Wi-Fi and Bluetooth connectivities), embedded sensors (e.g. gyroscope, accelerator) and decent storage memory. Typically, Acorn RISC Machine (ARM) CPU-based smartphones and tablets (e.g. Android OS, iOS devices) are the cost-efficient commercial products of integrated devices that can perform sensing tasks and also can interact with the cloud via the middle-edge facilities. Although the integrated devices may have constraint OS environment that reduces the flexibility of deploying virtualization platform on them; considering the swiftly evolved ARM CPUs and the embedded sensors embedded in the integrated devices, it is foreseeable that in near future, virtualization-based FEC will be available on the integrated devices. Overall, at this stage, a few platforms such as Apache Edgent (edgent.apache.org) or Termux (termux.com) are promising approaches towards realizing FEC on the integrated devices.

***IP gateway devices*** are also known as hubs, which act as the mediator between the constrained devices and the middle-edge devices. Commonly, because of the need for energy efficient wireless communication, many constraint devices do not operate in IP network, which usually requires the energy-intensive Wi-Fi (e.g. IEEE 802.11g/n/ac). Instead, the constraint devices are communicated using the protocols that consume less energy, such as Bluetooth Low Energy, IEEE 802.15.4 (e.g. ZigBee) or Z-Wave. Further, since the low energy communication protocols do not directly connect with the IP network, the system would use IP gateway devices to relay the communication messages between the constraint devices and the Internet gateway (e.g. routers). Hence, the backend cloud is capable to interact with the frontend constraint devices. In general, the Linux OS-based IP gateway devices such as Prota's hub (prota.info), Raspberry Pi or ASUS Tinker Board can easily host virtualization environment such as Docker Containers Engine. Hence, it is common to see that research projects [22] [23] [24] have been utilizing IP gateway devices as FEC node.

## 1.5 Business Models

While the common discussions of FEC are focusing on the advantages and applications, a fundamental question regards to how the business models of FEC will be like, has usually not been elaborated. Thereupon, here we discuss the three basic business models derived from the recent works [3] [10] [18] .

### *1.5.1 X as a Service*

Here, the 'X' of the X as a Service (XaaS) corresponds to infrastructure, platform, software, networking, cache or storage and many other types of resources mentioned in general cloud services. Specifically, XaaS providers of FEC allow their clients to pay to use the hardware equipment that supports SCANC mechanisms described in the previous section. Further, XaaS model does not limit to major business providers such as ISPs or the large cloud providers. Ideally, individuals and small businesses can also provide XaaS in the form of IndieFog [18] that is based on the popular Consumer as Provider (CaP) service provisioning model in multiple domains. For example, the MQL5 Cloud Network distributed computing project (cloud.mql5.com) utilizes Customer-Premises Equipments (CPEs) to perform various distributed computing tasks. Further, Fon (fon.com) utilizes CPEs to establish a global Wi-Fi network. These examples indicate that many individuals are willing to let application service providers pay to use their equipment for offering services.

## *1.5.2 Support Service*

The support service of FEC is similar to the software management support services in general information systems in which the clients who own the hardware equipment can pay the support service provider to provide them the corresponding software installation, configuration, and updates on the clients' equipment based on the requirements of the clients. Further, the clients may also pay for monthly or annual support services to the provider for assisting them with the maintenance and technical support. In general, support service providers offer their clients the highly customized solutions to achieve the optimal operation of their FEC-integrated systems. In general, a typical example of the support service provider is how Cisco provides the fog computing solution, in which the clients purchase Cisco's IOX-enabled equipment then pay the additional service fee to gain access to the software update and technical support for configuring their FEC environments. It is foreseeable that in near future, such a model will not be constrained to the single provider's hardware and software in which the support service provider will be decoupled from the hardware equipment vendors just like today's enterprise information systems support service providers such as RedHat, IBM or Microsoft.

## *1.5.3 Application Service*

Application service providers provide application solutions to help their clients in processing the data within or outside of the client's operation environments. For example, the recent Digital Twinning technologies create real-time virtualized 'twin' that clone the real-world behavior of a broad range of physical entities, from industrial facilities, equipment to the entire factory plane and the involved production lane and supply chains. Explicitly, such technology can provide the insight of the efficiency and performance towards optimizing and improving the industrial activities. Accordingly, an FEC application service provider can provide the Digital Twinning solution configured across all the involved entities at the edge networks in order to provide the analysis in an ultra-low latency manner (less than tens of milliseconds) towards helping the industrial system with the rapid reactions. Similarly, the FEC application service providers can also provide the service to local government in real-time traffic control system that assists the self-driving, connected vehicles. Further, IndieFog providers can also provide various application services to assist the Ambient Assisted Living (AAL) service providers in providing a certain edge analytics mechanisms for the clients of the AAL service providers. For example, an IndieFog provider who has installed Apache Edgent can offer the built-in stream data classification function as an application service for the mobile AAL clients in the close proximity.

## 1.6 Opportunities and Challenges

### *1.6.1 Out-of-Box Experience*

Industrial marketing research forecasts that the market value of FEC hardware components will reach $7,659 million by the year 2022 [10] , which indicates that more FEC-ready equipment such as routers, switches, IP gateway or hubs will be available in the market. Further, it is foreseeable that many of these products will feature with the Out-of-Box Experience (OOBE) in two forms—OOBE-based equipment and OOBE-based software.

**OOBE-based equipment** represents that the product vendors have integrated the FEC runtime platform with their products such as routers, switches or other gateway devices in which the consumers who purposed the equipment can easily configure and deploy FEC applications on the equipment via certain user interfaces, which is similar to the commercial router products that have graphical user interfaces for users to configure customised settings.

**OOBE-based software** is similar to the experience of Microsoft Windows in which the users who own FEC-compatible devices can purchase and install OOBE-based FEC software to their devices towards enabling FEC runtime environment and the SCANC mechanisms without any extra low-level configuration.

The OOBE-based FEC faces challenges in defining standardization for software and hardware. First, OOBE-based equipment raises a question to the vendors in what FEC platform and the related software packages should be included in their products? Second, OOBE-based software raises a question to the vendors regarding compatibility. Specifically, users may have devices in heterogeneous specification and processing units (e.g. x86, ARM etc.) in which the vendor may need to provide a version for each type of hardware. Moreover, developing and maintaining such an OOBE-based software can be extremely costly unless a corresponding common specification or standard for hardware exist.

## *1.6.2 Open Platforms*

At this stage, besides the commercial platforms such as Cisco IOX for fog computing, there is a few number of open platforms for supporting FEC. However, most of the platforms are in the early stage in which they have limited support in deployment. Below, we summarize the characteristics of each platform.

**OpenStack++** [25] is a framework developed by Carnegie Mellon University Pittsburgh for providing VM-based cloudlet platform on regular x86 computers for mobile application offloading. Explicitly, since the recent trend intends to apply lightweight virtualization technology-based FEC, OpenStack++ is less applicable to most use cases such as hosting FEC servers on routers or hubs. Further, it also indicates that the virtualization technology used in FEC is focusing more on containerization such as Docker Containers Engine.

**WSO2—IoT Server** (wso2.com/iot) is an extension of the popular open source enterprise service-oriented integration platform—WSO2 server that consists of certain IoT-related mechanisms such as connecting a broad range of common IoT devices (e.g. Arduino Uno, Raspberry Pi, Android OS devices, iOS devices, Windows 10 IoT Core devices etc.) with the cloud using standard protocols such as MQTT and XMPP. Further, WSO2—IoT server includes the embedded Siddhi 3.0 component that allows the system to deploy real-time stream processes in embedded devices. In other words, WSO2—IoT server provides the FEC computing capability on outer-edge devices.

**Apache Edgent** (edgent.apache.org), formerly known as Quarks, is an open source runtime platform contributed by IBM. Generally, the platform provides distributed stream data processing between cloud and edge devices. Specifically, the cloud-side supports most major open platforms in the stream data processing field such as Apache Spark, Apache Storm, Apache Flink and so forth. Further, at the outer-edge, Edgent supports common open operating systems such as Linux and Android OS. In summary, by utilizing Edgent, a system can dynamically migrate the stream data processing between cloud and edge, which ideally fulfils the need in most use cases that involve edge analytics.

Current open platforms lack capability in deploying and managing FEC across all the hierarchy layers of edge networks. However, it is likely due to the inflexibility of existing commercial devices in supporting the need for FEC mechanisms configuration. On the other hand, it also indicates an opportunity for product vendors to provide the enhanced devices that support FEC.

## *1.6.3 System Management*

Management of FEC involves the three basic life cycle phases—design, implementation, and adjustment.

***Design.*** The system administration team needs to identify where is the ideal location among the three edge tiers (i.e. inner-edge, middle-edge, outer-edge) for placing FEC servers [3] . Further, the administration team needs to develop or apply an ideal abstract modeling approach that can describe what types of resources the FEC servers are required with and how the FEC servers can interact with the system.

***Implementation.*** The administration team needs to consider the heterogeneity of FEC environments especially at the middle-edge and outer-edge where the nodes may have various hardware specification, communication protocols, and operating systems. Specifically, existing FEC equipment vendors (e.g. Cisco or Dell) may provide the isolated platform, which leads the implementation complex since the developers need to implement their FEC for each platform. Although there are a number of ongoing industrial-led open platform projects for FEC, the dependency requirement of each platform can still lead to a significant cost of time in the implementation.

***Adjustment.*** The FEC system needs to support runtime adjustment in which the system can schedule where and when to activate FEC functions in order to optimize the overall processes. For example, the system should have capabilities to dynamically deploy/terminate the runtime environment (e.g. VM or containers) and application methods on a feasible FEC node. Further, the system should be able to dynamically move the runtime environment or application methods from one FEC node to another based on the runtime context factors. Commonly, the required capabilities of adjustment phase raise challenges in how to support the reliability of software migration among FEC nodes and how to minimize the latency caused by such activities. In particular, dynamic code deployment and re-configuration at outdoor-based far-edge is highly challenging in terms of latency and reliability because the dynamic nature of wireless and mobile communication in which the signal interruption can cause the failure of code deployment [16].

## 1.7 Summary

Fog and edge computing (FEC) enhance the cloud-centric Internet of Things (CIoT) by extending the cloud computing model to the edge networks of IoT where the network intermediate nodes such as routers, switches, hubs and also the IoT devices are participating with the information processing and decision making towards improving security, cognition, agility, latency, and efficiency (abbreviated by SCALE).

This chapter provides an introductory overview of the state-of-the-art in FEC in terms of technical background, characteristics, deployment environment hierarchy, business models, opportunities and open challenges. Specifically, we have described the five fundamental advantages of FEC—SCALE, which is realized by the five mechanisms of FEC nodes—Storage, Compute, Acceleration, Networking and Control (SCANC). Further, to clarify the resource availability and their capabilities, this chapter has explained the three layers of FEC environment from the perspective of the central cloud in the core network. To enumerate, it consists of inner-edge with WAN providers, middle-edge with LAN and the frontline cellular networks, and outer-edge where the hubs and IoT devices are located at.

The capabilities of FEC will enable three types of business models known as X as a Service (XaaS), support service and application service. In summary, XaaS corresponds to the model that provides IaaS, PaaS, SaaS and S/CaaS (Storage or Caching as a Service), which are similar to existing cloud service models; support service corresponds to FEC software installation, configuration, and maintenance service that helps clients to set up their FEC on their own equipment; application service denotes the service providers cater the complete solution that serves FEC mechanism to clients without them to configure their own FEC system.

FEC brings new opportunities and also raises new challenges in development and operation. Specifically, development faces challenges in complexity and standardization, which potentially leads to the difficulty in system integration across different FEC providers and IoT end-points. Further, the operation challenge derives from the management cycle of FEC in terms of design, implementation, and adjustment. Explicitly, the heterogeneous network and entities involved in FEC led the challenges of FEC more complicated than the core network Internet-based cloud. On the other hand, the industry is aware of the challenges and has started a number of open platforms such as WSO2-IoT, Apache Edgent. Further, the recent Linux Foundation Project—EdgeX Foundry (edgexfoundry.org), which aims to provide a complete Software Development Kit for FEC, has shown the industrial interest in IoT is no longer satisfied with the connectivity between devices and cloud. Instead, the trend has moved from connected things to cognitive things in which the processes and decisions are performed as close to the physical objects as possible, even to the IoT devices themselves.